\documentclass[dvips]{article}
\usepackage{supertabular,lscape,epsfig}
\usepackage{amssymb}
\usepackage{amsmath}
\DeclareSymbolFont{ppa}{OT1}{ppl}{m}{it}
\DeclareMathSymbol{\vv}{\mathalpha}{ppa}{'166}

\begin{document}

\newcommand{\dd}{\,{\rm d}}
\newcommand{\ie}{{\it i.e.},\,}
\newcommand{\etal}{{\it et al.\ }}
\newcommand{\eg}{{\it e.g.},\,}
\newcommand{\cf}{{\it cf.\ }}
\newcommand{\vs}{{\it vs.\ }}
\newcommand{\zdot}{\makebox[0pt][l]{.}}
\newcommand{\up}[1]{\ifmmode^{\rm #1}\else$^{\rm #1}$\fi}
\newcommand{\dn}[1]{\ifmmode_{\rm #1}\else$_{\rm #1}$\fi}
\newcommand{\upd}{\up{d}}
\newcommand{\uph}{\up{h}}
\newcommand{\upm}{\up{m}}  
\newcommand{\ups}{\up{s}}
\newcommand{\arcd}{\ifmmode^{\circ}\else$^{\circ}$\fi}
\newcommand{\arcm}{\ifmmode{'}\else$'$\fi}
\newcommand{\arcs}{\ifmmode{''}\else$''$\fi}
\newcommand{\MS}{{\rm M}\ifmmode_{\odot}\else$_{\odot}$\fi}
\newcommand{\RS}{{\rm R}\ifmmode_{\odot}\else$_{\odot}$\fi}
\newcommand{\LS}{{\rm L}\ifmmode_{\odot}\else$_{\odot}$\fi}

\newcommand{\Abstract}[2]{{\footnotesize\begin{center}ABSTRACT\end{center}
\vspace{1mm}\par#1\par   
\noindent
{~}{\it #2}}}

\newcommand{\TabCap}[2]{\begin{center}\parbox[t]{#1}{\begin{center}
  \small {\spaceskip 2pt plus 1pt minus 1pt T a b l e}
  \refstepcounter{table}\thetable \\[2mm]
  \footnotesize #2 \end{center}}\end{center}}

\newcommand{\TableSep}[2]{\begin{table}[p]\vspace{#1}
\TabCap{#2}\end{table}}

\newcommand{\FigCap}[1]{\footnotesize\par\noindent Fig.\  %
  \refstepcounter{figure}\thefigure. #1\par}

\newcommand{\TableFont}{\footnotesize}
\newcommand{\TableFontIt}{\ttit}
\newcommand{\SetTableFont}[1]{\renewcommand{\TableFont}{#1}}

\newcommand{\MakeTable}[4]{\begin{table}[htb]\TabCap{#2}{#3}
  \begin{center} \TableFont \begin{tabular}{#1} #4
  \end{tabular}\end{center}\end{table}}

\newcommand{\MakeTableSep}[4]{\begin{table}[p]\TabCap{#2}{#3}
  \begin{center} \TableFont \begin{tabular}{#1} #4
  \end{tabular}\end{center}\end{table}}

\newenvironment{references}%
{
\footnotesize \frenchspacing
\renewcommand{\thesection}{}
\renewcommand{\in}{{\rm in }}
\renewcommand{\AA}{Astron.\ Astrophys.}
\newcommand{\AAS}{Astron.~Astrophys.~Suppl.~Ser.}
\newcommand{\ApJ}{Astrophys.\ J.}
\newcommand{\ApJS}{Astrophys.\ J.~Suppl.~Ser.}
\newcommand{\ApJL}{Astrophys.\ J.~Letters}
\newcommand{\AJ}{Astron.\ J.}
\newcommand{\IBVS}{IBVS}
\newcommand{\PASP}{P.A.S.P.}
\newcommand{\Acta}{Acta Astron.}
\newcommand{\MNRAS}{MNRAS}
\renewcommand{\and}{{\rm and }}
\section{{\rm REFERENCES}}
\sloppy \hyphenpenalty10000
\begin{list}{}{\leftmargin1cm\listparindent-1cm
\itemindent\listparindent\parsep0pt\itemsep0pt}}%
{\end{list}\vspace{2mm}}
 
\def\TYLDA{~}
\newlength{\DW}
\settowidth{\DW}{0}
\newcommand{\dw}{\hspace{\DW}}

\newcommand{\refitem}[5]{\item[]{#1} #2%
\def\REFARG{#3}\ifx\REFARG\TYLDA\else, {\it#3}\fi
\def\REFARG{#4}\ifx\REFARG\TYLDA\else, {\bf#4}\fi
\def\REFARG{#5}\ifx\REFARG\TYLDA\else, {#5}\fi.}

\newcommand{\Section}[1]{\section{#1}}
\newcommand{\Subsection}[1]{\subsection{#1}}
\newcommand{\Acknow}[1]{\par\vspace{5mm}{\bf Acknowledgements.} #1}
\pagestyle{myheadings}

\newfont{\bb}{ptmbi8t at 12pt}
\newcommand{\xrule}{\rule{0pt}{2.5ex}}  
\newcommand{\xxrule}{\rule[-1.8ex]{0pt}{4.5ex}}  
\def\thefootnote{\fnsymbol{footnote}}
\begin{center}

{\Large\bf
CURious Variables Experiment (CURVE).\\ CCD Photometry
and Variable Stars in the field of open cluster NGC 637}
\vskip1cm
{\bf
P.~~P~i~e~t~r~u~k~o~w~i~c~z$^1$,~~A.~~O~l~e~c~h$^1$,
~~M.~~W~i~\'s~n~i~e~w~s~k~i$^1$,\\
~~P.~~K~\c{e}~d~z~i~e~r~s~k~i$^2$,
~~K.~~M~u~l~a~r~c~z~y~k$^2$,
~~K.~~Z~{\l}~o~c~z~e~w~s~k~i$^{1,2}$\\
~~S.~~S~t~a~r~c~z~e~w~s~k~i$^1$,
~~and ~~K.~~S~z~a~r~u~g~a$^2$}
\vskip3mm
{
  $^1$Nicolaus Copernicus Astronomical Center,
     ul. Bartycka 18, 00-716 Warsaw, Poland\\
     e-mail: (pietruk,olech,mwisniew,star)@camk.edu.pl\\
  $^2$Warsaw University Observatory,
     Al. Ujazdowskie 4, 00-478 Warsaw, Poland\\
     e-mail: (pkedzier,kmularcz,kzlocz,kszaruga)@astrouw.edu.pl}
\end{center}

\Abstract{

We present $VI$ photometry for the open cluster NGC 637 which is located
in the Cassiopeia region. Morphology of cluster color-magnitude diagram
indicates that it is a young object with age of a few million years.
The apparent distance modulus of the cluster is $13.9<(m-M)_V<14.3$,
while reddening is $0.69<E(V-I)<0.73$. We estimated the heliocentric
distance as $2.6<d<3.3$ kpc. We also report the identification of two
variable stars in NGC 637. One of the variables is a non-radial pulsating
$\beta$ Cep-type star. Other one is a likely ellipsoidal variable,
however its pulsating nature can not be excluded.
}

{Hertzsprung-Russell (HR) and C-M diagrams --
Stars: variables : beta Cep -- open clusters and associations:
individual: NGC 637}

\Section{Introduction}

CURious Variables Experiment (CURVE) is a long-term project focused
on observations of open clusters, globular clusters and cataclysmic
variable stars in the northern hemisphere (Olech \etal 2003a, Olech
\etal 2003b). In stellar clusters we principally search for variable
objects. However, our data also allow to estimate basic parameters
of observed clusters, such as distances and ages.

NGC 637, also known as C 0139+637 or OCl 329, is a not very abundant
open cluster situated in the Cassiopeia region of the Perseus spiral arm
of our Galaxy ($l$=128\zdot\arcd55, $b$=+1\zdot\arcd73). The first photometric
studies of this cluster were done by Grubissich (1975) in $RGU$ system.
He estimated reddening and distance to the cluster. Later,
Huestamendia \etal (1991) published photoelectric $UBV$ observations
and assessed age of the cluster as 15 Myr. NGC 637 was also observed by Phelps
and Janes (1994) as one of 23 open clusters to explore the star formation
history of the Cassiopeia region. They found that the previous
authors had overestimated the age of the cluster and suggested NGC 637
to be 0-4 Myr old.

This paper presents CCD $VI$ photometry and results of a year-long
monitoring of NGC 637 in search for variable objects.
We also give characteristic parameters of the cluster.

\Section{Observations and Reductions}

Observations of the open cluster NGC 637 were made during 52 nights
between January 29, 2004 and December 11, 2004 at the Ostrowik
station of the Warsaw University Observatory. The data were collected
using the 60-cm Cassegrain telescope equipped with a Tektronics
TK512CB back-illuminated CCD camera. The scale of the camera was
$0\zdot\arcs76$/pixel providing a $6\zdot\arcm5 \times 6\zdot\arcm5$
field of view. The full description of the telescope and camera
was given by Udalski and Pych (1992).

We monitored the cluster using standard Johnson-Cousins $VI$ filters
and also in "white light". The exposure times were from 120 to 240 seconds.
In total, we obtained 124, 234 and 591 images in $V$, $I$ and white light,
respectively. The median seeing was 4\zdot\arcs03, 4\zdot\arcs06
and 3\zdot\arcs90, respectively. The map of the observed region is shown
in Fig.~1.

\begin{figure}[htb]
\centerline{\includegraphics[height=100mm,width=100mm]{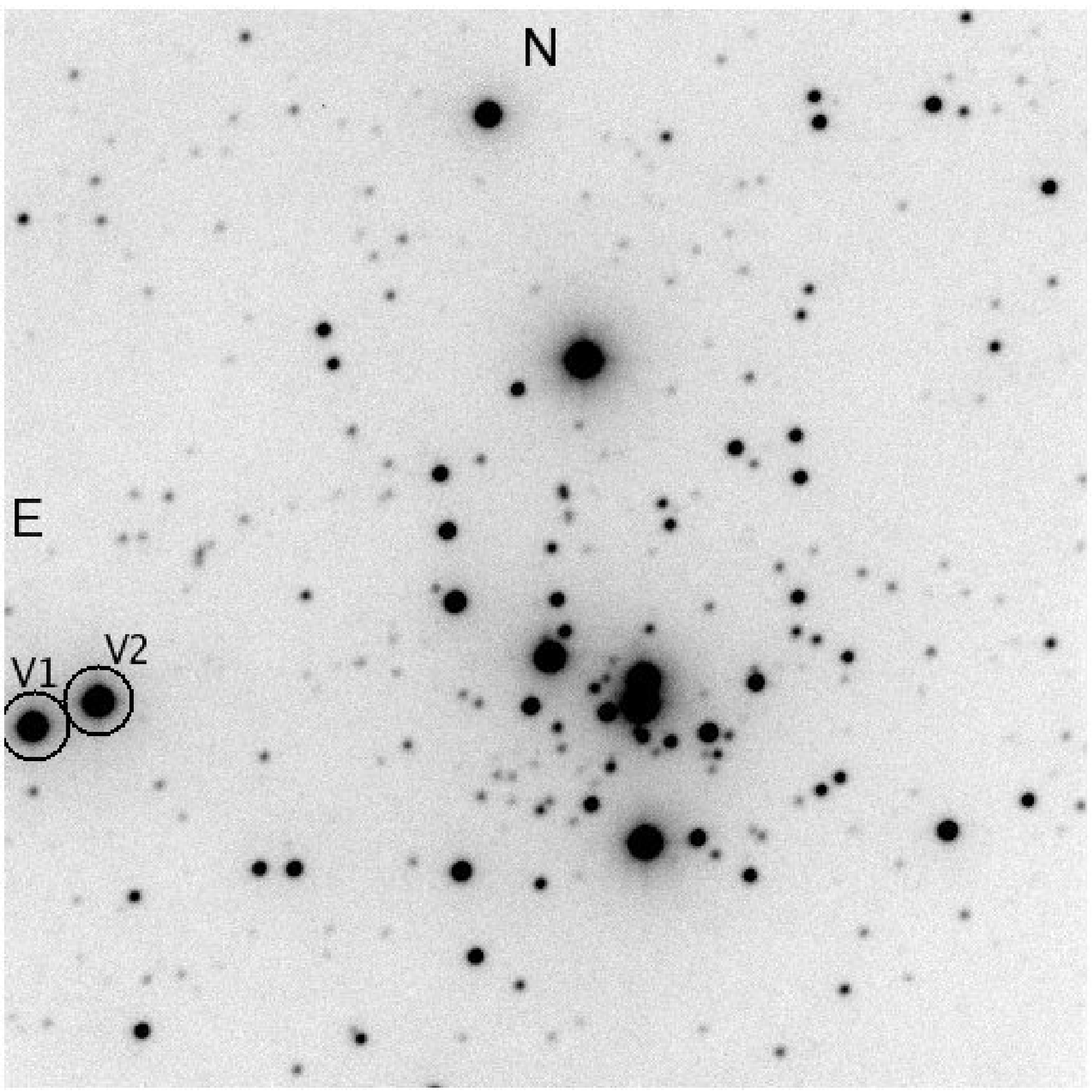}}
\FigCap{
Map of NGC 637 based on a white light image taken with 180 sec exposure.
The field of view is about $6.5 \times 6.5$ arcmin$^2$. Detected variable
stars are marked by circles.
}
\end{figure}

All images were de-biased, dark current subtracted and flat-fielded
using the IRAF \footnote{IRAF is distributed by the National Optical
Astronomy Observatory, which is operated by the Association of
Universities for Research in Astronomy, Inc., under a cooperative
agreement with the National Science Foundation.} package.
For the white light images the photometry was extracted with the help
of a slightly modified ISIS-2.1 image subtraction package (Alard
and Lupton 1998). Our procedure followed that described in detail
by Mochejska \etal (2002). A reference frame was constructed by
combining 23 individual images taken during dark time on the 
night of 2004 Sep 18/19. The seeing for the resultant reference image
was FWHM=2.88 arcsec. Profile photometry for the reference frame
was extracted with DAOPHOT/ALLSTAR (Stetson 1987).
These measurements were used to transform the light curves
from differential flux units into instrumental magnitudes.

For the $V$ and $I$ band images we derived classical profile
photometry. In each filter we selected one image as a reference
and calculated the median value of the magnitude offset for each
individual frame in respect to the reference image. Subsequently,
we constructed light curves for all stars retained on lists corresponding
to the reference image.

All light curves were analyzed in a search for variable stars.
This was performed with the TATRY code using multiharmonic
periodogram of Schwarzenberg-Czerny (1996). Periodograms were
calculated for periods ranging from 0.05 to 500 days,
and the light curves examined by eye.

We perform an elimination of some potentially poor measurements
in lists of stars extracted from $V$ and $I$ reference frames.
A few objects with unusually large errors of photometry, \ie 
with large values of CHI and SHARP parameters, returned
by DAOPHOT, were rejected. Fig.~2 displays the $rms$ deviation
as a function of magnitude in $V$ and $I$ bands for 181 and 284
stars, respectively.

\begin{figure}[htb]
\centerline{\includegraphics[height=80mm,width=120mm]{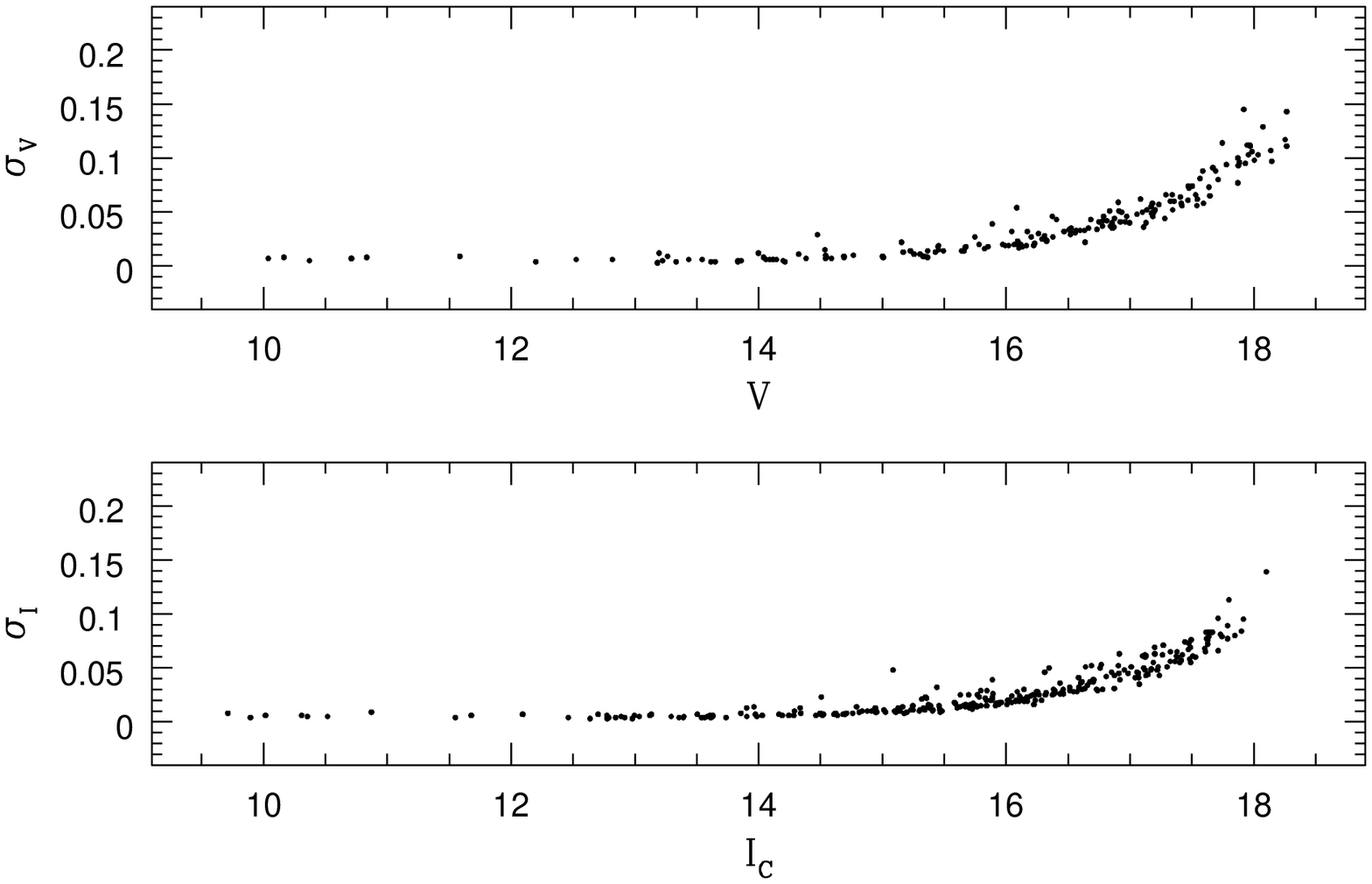}}
\FigCap{
Plots of $rms$ magnitude as a function of $V$ (top) and $I$ (bottom)
magnitude for stars in the region of NGC 637.
}
\end{figure}

Finally, magnitudes of all stars were transformed from our instrumental
system to photometry obtained by Henden (2001). Basing on the brightest
common stars we calculated coefficients of transformation. Equations
are the following:
$$
i=I-0.122(V-I)+1.017\\
\eqno(1)
$$
$$
v-i=1.043(V-I)-0.162.
\eqno(2)
$$
Appropriate aperture corrections were derived proceeding the transformation.

\Section{Results}

\Subsection{Density profile}

The center of NGC 637 was found by calculating the density center
for stars inside a circle of radius of 70 pixels, using an
iterative procedure similar to that described by Mateo \& Hodge (1986).
In Fig.~3 we show the density profile for all stars brighter
than $I=17.3$. The average stellar density was calculated in successive
13.68 arcsec (18 pixels) wide annuli around the cluster center. The resulting
density profile is rather noisy due to small number statistics.
The smooth solid line represents a fit by the King (1962) profile:
$$
f(r) = \frac{f_0}{1+(r/r_c)^2} + f_b
\eqno(3)
$$
where $f_0$ is the central density, $r_c$ is the radius of the cluster
core and $f_b$ is the background density. For NGC 637 we found as follows:
$f_0=0.010 \pm 0.003$ stars per arcsec$^2$,
$r_c=20 \pm 4$ arcsec and $f_b=0.0013 \pm 0.0001$ stars/arcsec$^2$.
The cluster appears to be rather loose and has a small clumpy center.
It is likely that some stars of the cluster are located outside our field.
This may be due to the fact that {\it gross} stars from entire our field
have consistent proper motions (\eg from a survey by Monet \etal 2003).

\begin{figure}[htb]
\centerline{\includegraphics[height=65mm,width=120mm]{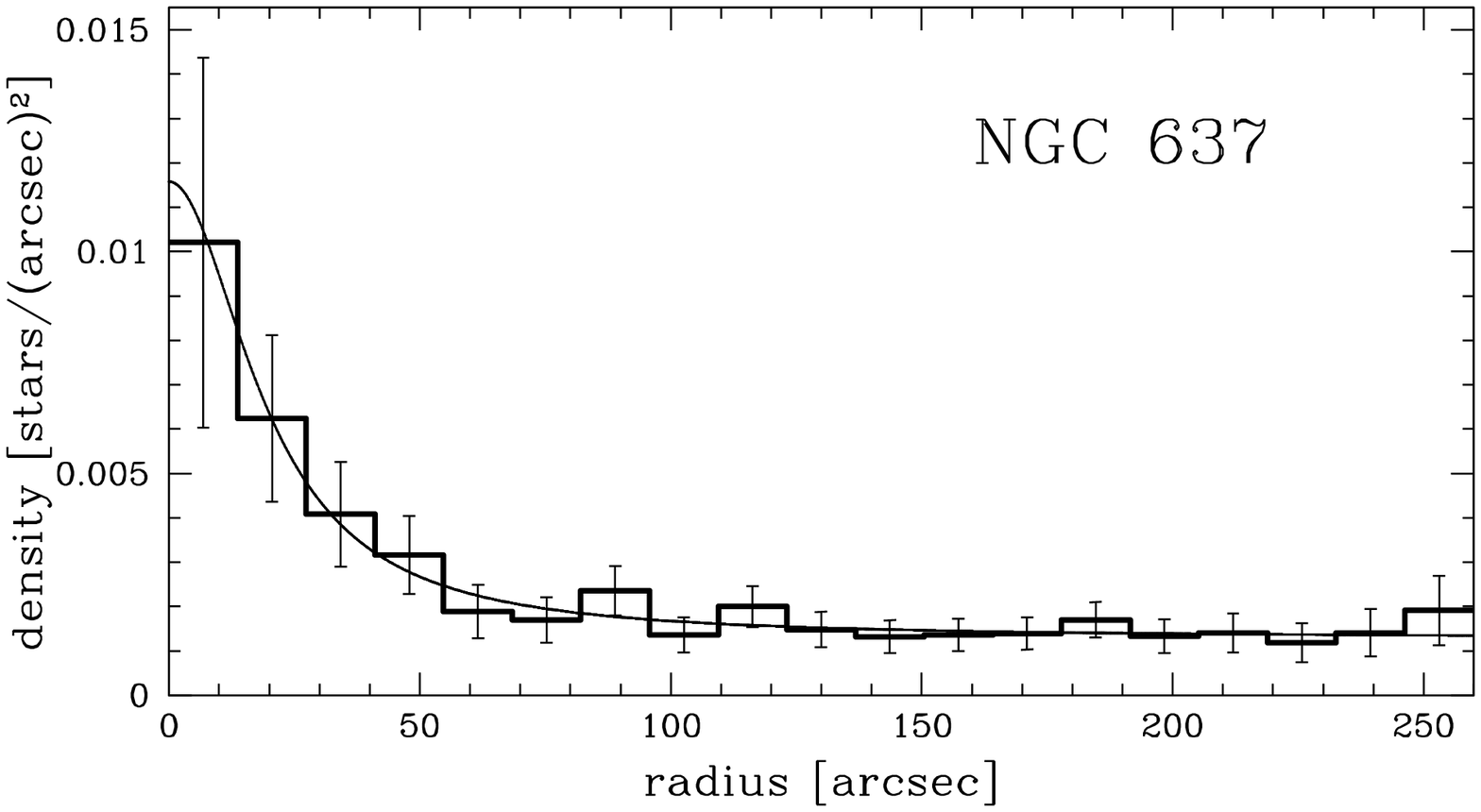}}
\FigCap{
Surface density distribution of stars with $I<17.3$ from the NGC 637
field. The smooth solid line represents the King (1962) profile
fitted to the histogram.
}
\end{figure}

\Subsection{Color-magnitude diagram}

Fig.~4 presents color-magnitude diagram (CMD) for the observed field with
the cluster NGC 637. The morphology of CMD is typical of young open cluster.
A gap in the upper main sequence, noted by Huestamendia \etal (1991),
is also present in our data. This confirms that it does not result
from incompleteness of the data.

\begin{figure}[htb]
\centerline{\includegraphics[height=120mm,width=120mm]{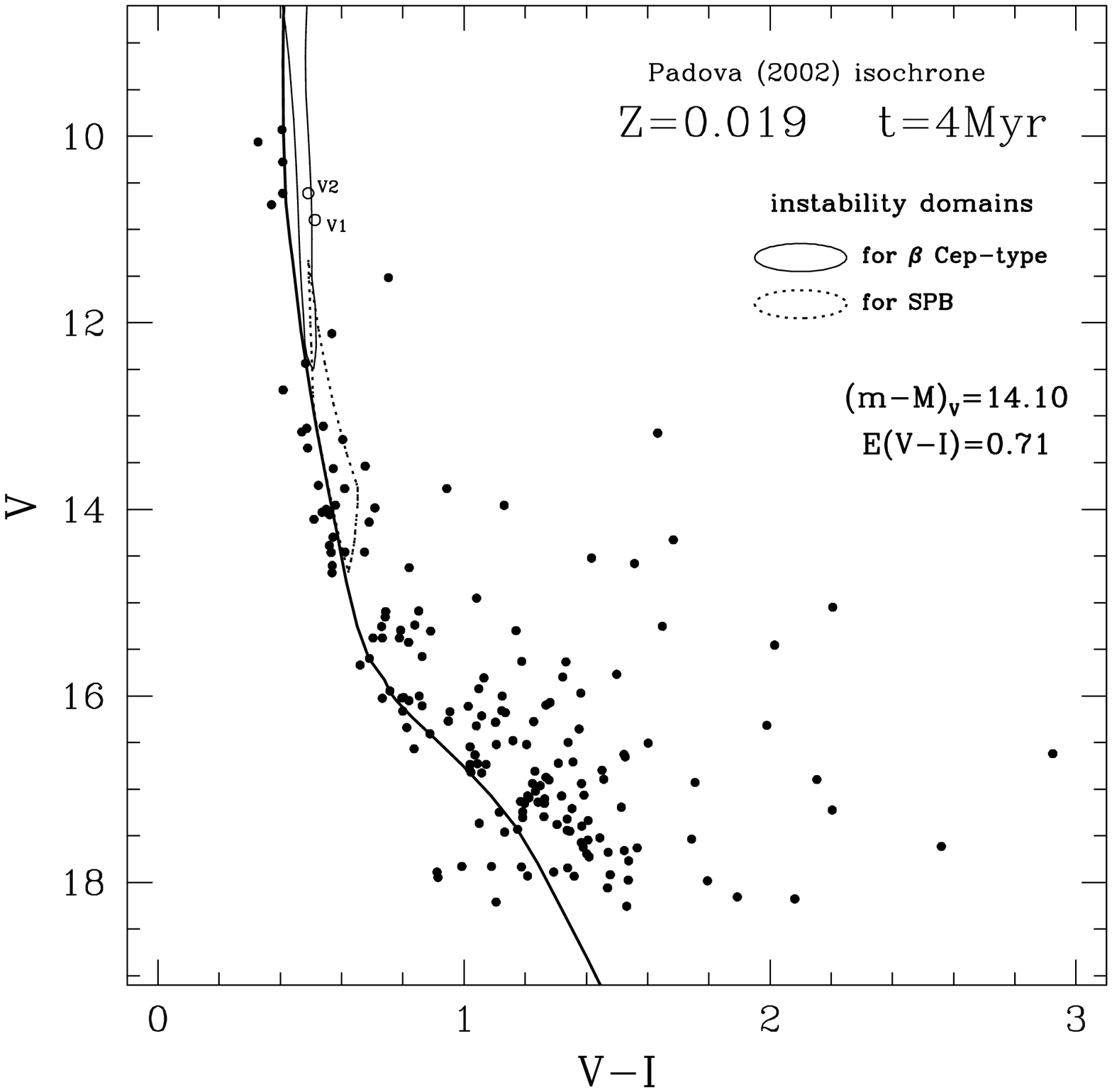}}
\FigCap{
$V/V-I$ diagram for NGC 637 with a superimposed isochrone
and instability domains in the upper main sequence.
Positions of two detected variables are denoted by small circles.
}
\end{figure}

Using theoretical isochrones published by Girardi \etal (2002) we are able to
estimate the basic parameters of NGC 637. We fit an isochrone with solar chemical
composition $(Z,Y)=(0.019,0.273)$ and the youngest available age, 4.0 Myr.
The shape of the main sequence is well reproduced for stars with $V<17$ mag.
We derived, by shifting the isochrone, the apparent distance modulus
of $(m-M)_V=14.10$, and the reddening of $E(V-I)=0.71$. Assuming
$E(V-I)=1.28E(B-V)$ we get $E(B-V)=0.55$. This value is significantly lower
than the upper limit of reddening $E(B-V)=1.21$ for $(l,b)=(128.55,+1.73)$,
extracted from the maps of Schlegel \etal (1998). We assume the uncertainty
of the distance modulus as 0.2 mag, and the uncertainty of the reddening
$E(V-I)$ as 0.02 mag. Adopting $R_V=3.2$ we find the minimum value
of the heliocentric distance to NGC 637 as $d_{min}=2.6$ kpc and the maximum
value as $d_{max}=3.3$ kpc. These numbers are consistent with the distance
$D=2884$ pc given by Phelps and Janes (1994). On the other hand, our value
of the reddening $E(B-V)$ differs from their value of 0.65 mag.

\Subsection{Variable stars}

The search for variability led to the detection of two variables.
They are located close to the edge of our field and are among the brightest
stars (see the map in Fig.~1). Table 1 lists basic data for the
new variables. Fig.~4 displays their location on the $V/V-I$ color-magnitude
diagram with superimposed instability domains in the upper main sequence.
The domains were computed using the OPAL opacities for the hydrogen content
$X=0.70$ and metallicity $Z=0.019$ (same as in Pamyatnykh 1999),
and transformed to the observational values using numerical relations
from Flower (1996).

\begin{table}
\centering
\caption{\small Equatorial coordinates and photometric data for the NGC 637 variables}
{\footnotesize
\begin{tabular}{lccccccccc}
\hline
Name & RA(2000.0) & Dec(2000.0) & $<V>$ & $<I>$ & $\Delta V$ & $\Delta I$ & $P_0$ & $P_1$ & Type \\
\hline
V1 & 01\uph43\upm35\zdot\ups57 & +64\arcd02\arcm07\zdot\arcs0 & 11.38 & 10.87 & 0.14 & 0.16 & 0.189991(3) & 0.18917(1) & $\beta$~Cep \\
V2 & 01\uph43\upm32\zdot\ups18 & +64\arcd02\arcm15\zdot\arcs3 & 11.12 & 10.63 & 0.04 & 0.06 & 0.74316(8)  &     -      & ell ? \\
\hline
\end{tabular}}
\end{table}

\begin{figure}[htb]
\centerline{\includegraphics[height=120mm,width=120mm]{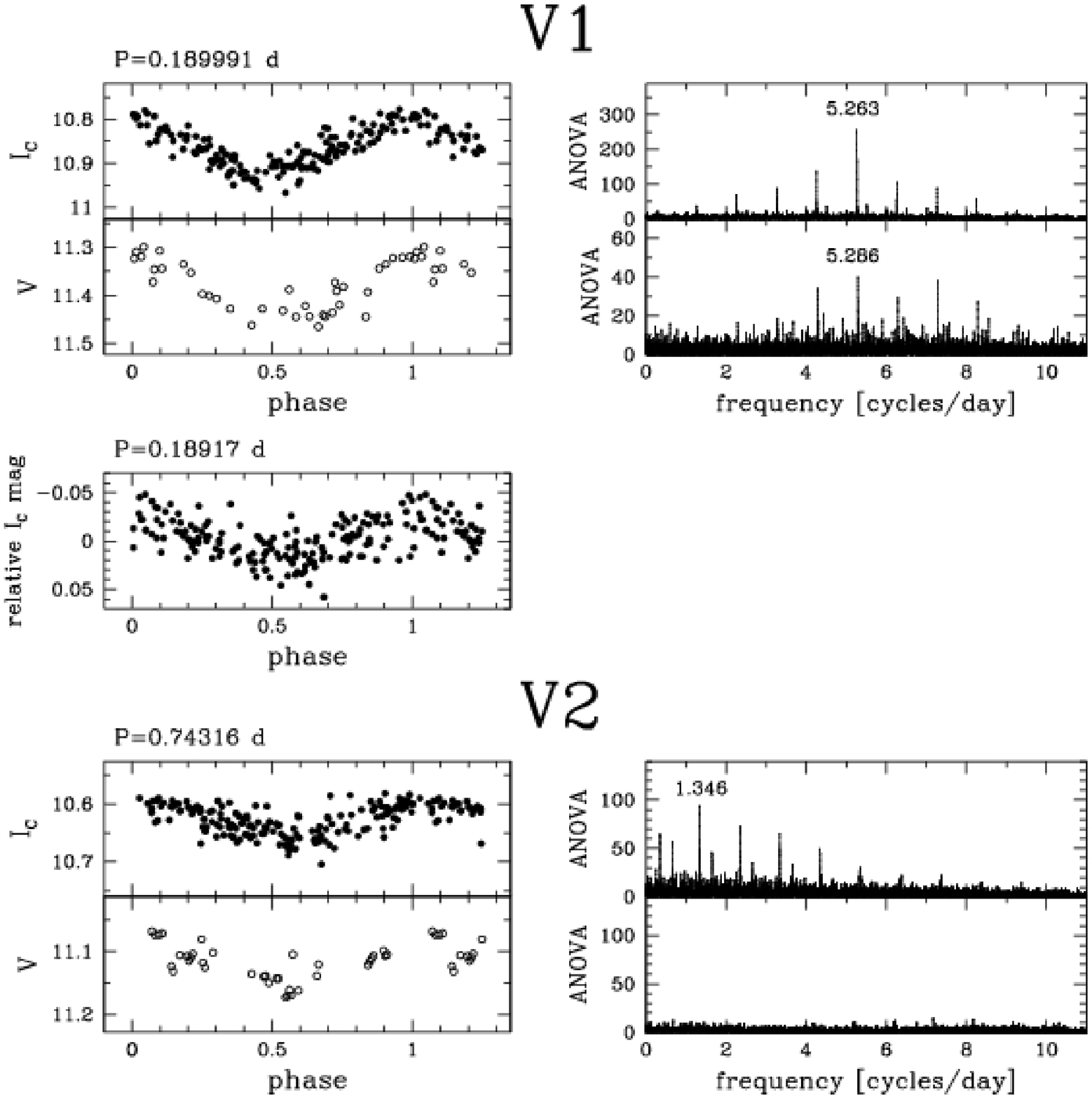}}
\FigCap{
Phased light curves and power spectra of two variables detected
in the field of NGC 637.
}
\end{figure}

Fig. 5 presents the $V$ and $I$ light curves of newly identified
variables. The primary period of V1 is $P=0.189991$ days. Its value,
together with the position of the star on the CMD and amplitude of the
light modulations, clearly indicates that V1 belongs to the $\beta$
Cep-type variables.

The right panels of Fig. 5 show the ANOVA power spectra of the light
curves (Schwarzenberg-Czerny 1996). The original light curve of V1 was
prewhitened with main frequency and its harmonic and the resulting ANOVA
statistics returned another clear periodicity of $P=0.18917$ days. The
difference between two identified frequencies is 0.023 c/d, which is
over ten times larger than our frequency determination errors, and
indicates that both periods are real. Such multiperiodic behavior caused
by non-radial pulsations is common among $\beta$ Cep-type stars
(\eg Shobbrook \etal 2006, Schrijvers \etal 2004, Schrijvers and Telting
2002).

The variable V2 is even brighter than V1, being one of the brightest
stars in the cluster. Its period is $P=0.74316$ days, which is typical
for Slowly Pulsating B stars (SPB stars) but the variable seems to
be too bright to belong to this group. The dotted line plotted in CMD
shown in Fig. 4 denotes instability domain of SPB stars and clearly
demonstrates that V2 is located about 1 mag over the tip of the SPB
domain.

Another possibility is that V2 is in fact an ellipsoidal variable with
orbital period of 1.48632 days. It is justified by small amplitude of
brightness modulations and its nearly sinusoidal shape. In our opinion
this is the more likely explanation.

Proper motions of the two detected variables (from \eg Kislyuk \etal 1999,
Hog \etal 2000, Kharchenko 2001, Monet \etal 2003) suggests their
membership to NGC 637.

In their paper, Huestamendia \etal (1991), basing on merely six
observational points, remarked that their object \#18 is variable.
Our data do not show any variability of the object.

It is also worth noting that we searched for dwarf novae outburst
in the field of NGC 637. We used the procedure described in detail
by Kaluzny \etal (2005) and we applied it here to white light data.
Result of this search is negative, which is not surprising in such
a young cluster.

\Section{Summary}

We have presented the CCD $VI$ photometry and results of searches
for variable stars in the open cluster NGC 637. The analysis
of the derived color-magnitude diagram allowed us to assess
some basic parameters. The apparent distance modulus for the cluster
is $13.9<(m-M)_V<14.3$, the reddening is $0.69<E(V-I)<0.73$,
and the estimated heliocentric distance of $2.6<d<3.3$kpc.
The morphology of the CMD indicates that it is a young object
with age of a few million years. Based on the comparison
of the CMD with the theoretical isochrones, we were also not
able to firmly establish the metallicity of the cluster.
Spectroscopic observations would help to clarify it.

The search for variability led to discovery of two variable stars
in NGC 637. One of them, V1, is a bona fide $\beta$ Cep-type variable.
It has two close periodicities what clearly indicates non-radial
pulsations. The light curve of object V2 has nearly sinusoidal shape
and small amplitude and therefore we classified it as an ellipsoidal
variable. However, one can not excluded a possibility that
V2 is a pulsating star.

\Acknow{

The authors would like to thank Drs. A. Pamyatnykh and W. Dziembowski
for helpful comments on detected variables and Dr. W. Pych
for providing some useful software which was used in the analysis.
This work was supported by Polish KBN grant number
1~P03D~006~27 to A. Olech.
}

\end{document}